\documentclass[]{emulateapj}

\usepackage{xspace}
\def\andix{And\,IX\xspace}

\shorttitle{And IX:  Properties of the Smallest M\,31 Dwarf Satellite Galaxy}
\shortauthors{Harbeck et al.}

\begin{document}


\title{Andromeda IX: Properties of the Smallest M\,31 Dwarf Satellite Galaxy}


\author{Daniel Harbeck\altaffilmark{1,2}, John
  S. Gallagher\altaffilmark{1}, Eva K. Grebel\altaffilmark{3}, Andreas
  Koch\altaffilmark{3}, and Daniel B.  Zucker\altaffilmark{4}}

\altaffiltext{1}{University of Wisconsin, Astronomy Department, 475
  N. Charter St., Madison, WI 53706; harbeck@astro.wisc.edu, 
  jsg@astro.wisc.edu}

\altaffiltext{2}{University of California, Space Sciences Laboratory,
  7 Gauss Way, Berkeley, CA 940720; harbeck@ssl.berkeley.edu}

\altaffiltext{3}{Astronomical Institute of the University of Basel,
  Venusstrasse 7, 4102 Binningen, Switzerland; grebel@astro.unibas.ch,
   koch@astro.unibas.ch}

\altaffiltext{4}{Max-Planck Insitut f\"ur Astronomy, K\"onigstuhl 17,
  69117 Heidelberg, Germany; zucker@mpia.de}



\begin{abstract}
 We present WIYN observations of the recently discovered And IX dwarf
 spheroidal (dSph) satellite galaxy of M\,31. Our data, obtained at a
 natural seeing of $0.5''$ and just reaching the horizontal branch
 level, confirm \andix as a dSph galaxy with a distance similar to
 M\,31. A survey for carbon stars shows no evidence for an
 intermediate-age (1--10\,Gyr) stellar population in \andix. From the
 red giant branch we estimate a metallicity of roughly
 $-2$\,dex. Combined with the tip of the red giant branch luminosity,
 this results in a distance of $735$~kpc, placing \andix approximately
 $45$\,kpc from M\,31. This faint dSph follows the relations between
 luminosity and metallicity, and luminosity and surface brightness
 defined by other Local Group dSph galaxies. The core and tidal radii
 are found to be $1.35'$ and $5.9'$, respectively. We conclude that
 \andix -- despite its low luminosity -- might be an ordinary Local
 Group dSph and discuss implications for its formation from a once
 more massive, but stripped progenitor or from an intrinsically
 low-mass seed.

\end{abstract}




\section{Introduction}\label{Sect_Intro}

The galaxy census of the Local Group (LG) remains incomplete.  Only
five years ago four new LG members were discovered, increasing the
total census to 36.  All of the new galaxies turned out to be dwarf
spheroidal (dSph) galaxies, the faintest, apparently least massive,
most diffuse and gas-deficient galaxy type known (for a detailed
discussion of dSph properties, see \citealt{grebel2003}).  The new
discoveries add to the faint end of the galaxy luminosity function of
the LG \citep{grebel1999}, but are far too few to compensate for the
observed lack of low-mass dwarf galaxies as compared to the
cosmologically predicted high number of low-mass dark matter halos
(e.g., \citealt{moore1999}).  The new dwarfs were detected on POSS-I,
POSS-II, and ESO/SERC EJ plates.  Thorough visual all-sky searches of
these photographic data have since been completed without revealing
additional LG members (e.g., \citealt{kara2000};
\citealt{whiting2002}).  This suggested that only very few future
detections were to be expected, possibly in regions of reduced
sensitivity in the photographic plates or in areas of high dust
obscuration, unless there is a population of dwarf galaxies with
fainter surface brightness than currently known.

Recently a new candidate LG member was found in Sloan Digital Sky
Survey (SDSS) data \citep{zucker2004}.  It was named Andromeda IX
since it seems to be a dSph companion of M31. Zucker et al.\ report
the detection of an apparent metal-poor upper red giant branch (RGB)
traced to approximately 2\,mag below its tip, and estimate a distance
of $\sim 790$\,kpc.  With an absolute magnitude of only $M_V \sim
-8.3$ mag and a central surface brightness of $\mu_{V,0} \sim
26.8$\,mag\,arcsec$^{-2}$ \citep{zucker2004}, \andix would then be the
lowest luminosity galaxy known.  If confirmed, this raises a number of
important questions: Is there a vast population of these extremely
low-luminosity dwarfs in the LG and elsewhere still awaiting
detection, which may now become feasible with sensitive and
homogeneous imaging surveys?  What are the properties of these
objects?  Are they simply the faint extension of the known dSph
population?  Do they show evidence for intrinsic population gradients
as seen in the old populations of ``normal'' dSphs
\citep{harbeck2001}, indicative of extended star formation episodes?
Do these objects share a common epoch of ancient star formation as
expected from reionization scenarios (see, e.g.,
\citealt{grebel2004})?  How faint are the least massive dark matter
halos that still contain baryonic mass?  Does the galaxy luminosity
function rise sharply at its faint end?  The first step toward a
better understanding of these questions is an in-depth study of the
unusual object \andix.  In this Letter, we present deep ground-based
photometry of \andix to verify its status as a LG member and M31
companion and to uncover its properties.

\section{Observations and Data Reduction}

 \andix was observed during the photometric night of 2004 September 22
 with the WIYN 3.5-m telescope\footnote{The WIYN Observatory is a
   joint facility of the University of Wisconsin-Madison, Indiana
   University, Yale University, and the National Optical Astronomy
   Observatories.}.  We used the OPTIC detector, which is a mosaic
 camera consisting of two orthogonal transfer CCDs with a total field
 of view of $9.5'\times9.5'$. The two CCDs have an increased red
 sensitivity. During our observations of \andix the seeing was always
 better than $0.5''$.  With \andix centered on chip \#2, which is free
 of cosmetic defects, we obtained two 500\,sec images in the V band,
 one 500\,sec exposure in the I band, and 3$\times$500\,sec exposures
 in two narrow-band filters centered on the near-infrared TiO and CN
 bands.  These narrow-band filters provide a reliable method to
 identify carbon stars (see., e.g., \citealt{cook1986,
   harbeck2004}). The data were reduced in a standard way including
 overscan, bias, and flat field corrections. If multiple images in one
 filter were available, we combined these images with cosmic ray
 rejection enabled.

 From the reduced images we obtained stellar photometry using the
 DAOPHOT implementation under IRAF\footnote{IRAF is distributed by the
   National Optical Astronomy Observatories, which are operated by the
   Association of Universities for Research in Astronomy, Inc., under
   cooperative agreement with the National Science Foundation.}. We
 additionally obtained observations of five standard star fields
 \citep{landolt1992} in the V and I filters from which we derived the
 photometric solution. Then we transformed our instrumental magnitudes
 to the Landolt system, including aperture-PSF shift, photometric
 zeropoint, and first order color and airmass terms.

\section{Results}

\subsection{Color-magnitude diagram, distance, and metallicity}

\begin{figure}
\plotone{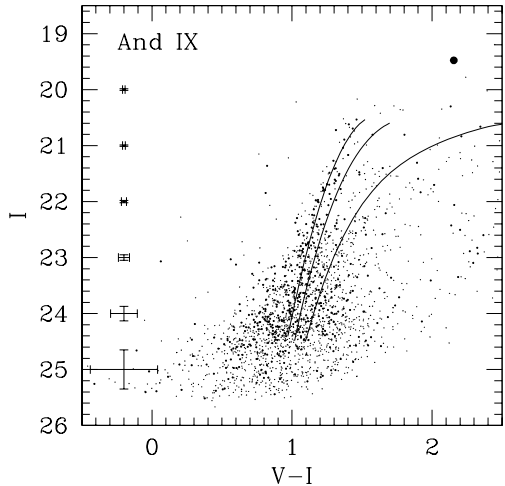}
\caption{Color magnitude diagram of \andix based on WIYN/OPTIC
  observations. Stars within 1.25' of \andix's center are plotted with
  thick dots. Representative error bars are shown at the left side. We
  overplot red giant branch fiducials from \citet{fiducials90} for
  M15, M2, and 47\,Tuc, which have metallicities of
  [Fe/H]=$-2.26$\,dex, $-1.62$\,dex, and $-0.76$\,dex,
  respectively. The fiducials have been reddened and shifted in
  distance as appropriate for \andix. The large dot represents the
  location of the carbon star candidate.
 \label{fig_cmd}}
\end{figure}

We present the resulting color-magnitude diagram (CMD) of \andix in
Fig.\,\ref{fig_cmd}. We plot only objects that are likely point
sources according to their DAOPHOT sharpness and chi parameters.
Owing to the excellent seeing and to the longer exposure times, our
data go roughly 2.5 mag deeper than the data available in the
discovery paper \citep{zucker2004}.  Stars within $1.25'$ of the
galaxy center are marked with larger dots in the CMD. While the total
stellar density is low compared to the background contamination, a
steep red giant branch is clearly visible, although this RGB is only
sparsely populated.  At a luminosity of $I\sim24.5$, approximately
1\,mag above our detection limit, a possible horizontal branch and/or
an old red clump become apparent. There is no indication of a luminous
blue main sequence.  Thus our observations confirm \andix as a nearby
dwarf galaxy; confusion with a background galaxy (as happened with And
IV, \citealt{ferguson2000}) can be ruled out.

The CMD shows what appears to be a second, but considerably more
metal-rich RGB that extends beyond $(V-I) > 2.5$~mag.  Stars along
this wide feature do not show any obvious concentration toward the
center of And\,IX.  Considering that the projected distance to M31's
center is only $\sim$30~kpc, it seems likely that many of the
``background'' stars in the CMD belong in fact to the extended outer
disk or to the halo of M31.  In contrast to the Galactic halo, M31's
halo is considerably more metal-rich with a metallicity peak at [Fe/H]
$\sim -0.5$ dex \citep{Durrell2004}, consistent with the location of
the red stars in our CMD.

\begin{figure}
\plotone{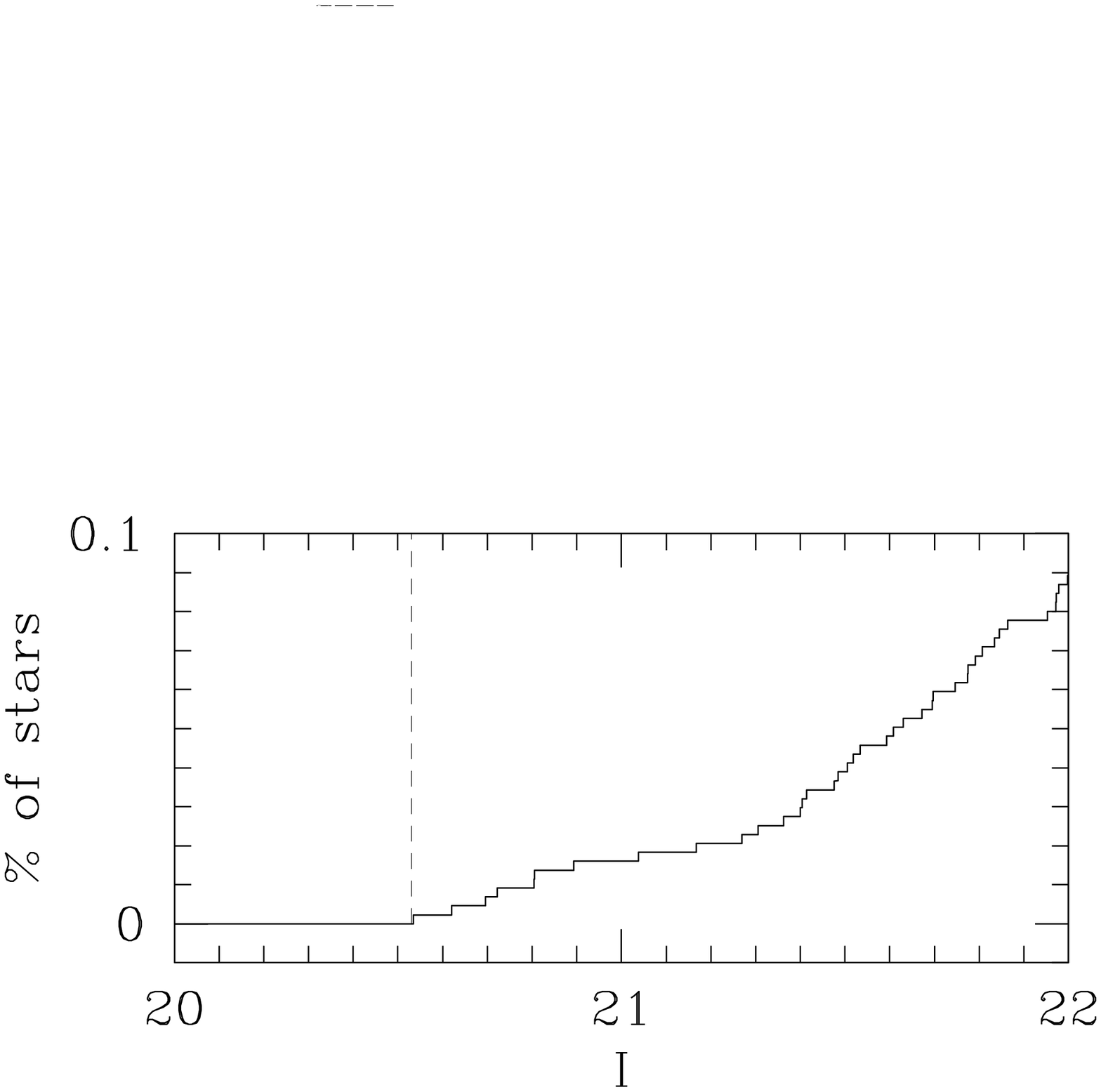}
\caption{Luminosity function of stars within a radial distance of
  $1.25$' and $(B-V) \le 2$. The tip of the RGB is dominated by a
  single star at $I=20.5$~mag, and we estimate a uncertainty of this
  luminosity of order of $0.1$~mag due to the spare sampling of the
  RGB.}
\label{fig_lumfunc}
\end{figure}

We use the I-band luminosity function for stars with $(V-I)\le2$\,mag
and with a radial distance $\le 1.25$' (see Fig.~\ref{fig_lumfunc}) to
measure the luminosity of the tip of the RGB (TRGB) to $I_{TRGB}
=20.53$\,mag (same procedure as in \citealt{greguh1999}). Due to the
sparse sampling of the RGB the single brightest RGB star in \andix's
central region determines the rise of the luminosity function (compare
with Fig~\ref{fig_cmd}). Given the uncertainties in the photometric
calibration and the very few stars near the RGB tip we assume a
conservative uncertainty of $\Delta I_{TRGB}=0.1$\,mag. Confusion
problems due to M31's field population, which is at roughly the same
distance, aggravate the distance determination for \andix.  However,
since there is a clear increase of RGB candidate stars as well as a
possible increase of red horizontal branch stars at the location of
\andix, we are confident about our TRGB distance estimate.  We use an
absolute I band luminosity of $-3.95$\,mag for the TRGB assuming a
metallicity for \andix of [Fe/H]=-2~dex (see next paragraph) following
\citet{bellazzini2004}.  Adopting a foreground reddening of
E(B-V)$=0.076$ and $A_I=1.94 \cdot E(B-V)$ from \citet{schlegel1998},
we arrive at a distance modulus for And\,IX of $20.53+3.95-1.94 \cdot
0.076=24.33\pm0.1$\,mag.  Our derived distance agrees within the
uncertainties with the value of $24.48$\,mag derived by
\citet{zucker2004} and with McConnachie et al.'s (2004) result of
$24.42$ mag, both of which are based on shallower data. While we
derive the same TRGB luminosity as \citet{mcconnachie2004} within
0.03~mag, the different distance modulus is due to a different adopted
metallicity (our -2.0 dex vs. their -1.5 dex), which alters the
absolute TRGB luminosity. \andix's distance modulus corresponds to a
distance of $735\pm15$\,kpc. If we adopt the Cepheid distance of 770
kpc to M31 \citep{madore1991}, this yields a deprojected distance of
$45\pm15$\,kpc between \andix and M\,31's center; this confirms that
\andix is not only a plausible M\,31 satellite, but also may be its
closest dSph companion.

\subsection{Metallicity}

In the CMD in Fig.~\ref{fig_cmd} we plot RGB mean ridge lines of the
Milky Way globular clusters M\,15, M\,2, and 47\,Tuc from
\citet{fiducials90}. These globular clusters have metallicities of
      [Fe/H]=$-2.26$\,dex, $-1.62$\,dex, and $-0.76$\,dex respectively
      \citep{harris1996}. The fiducials have been reddened adopting a
      foreground reddening of E(B$-$V)=0.076 and shifted according to
      \andix's distance modulus. The comparison of its RGB with the
      fiducials suggests that \andix has a low metallicity in the
      range of $-2$\,dex, or even lower. This metallicity estimate,
      however, is valid for stellar ages of order of 10~Gyr or older
      only, an assumption that is justified given the lack of
      indication of younger stellar populations in \andix.
      Independent qualitative support for the And\,IX distance derived
      in Section 3.1 comes from the good match between the observed
      RGB and the fiducials (cf. Fig.~\ref{fig_cmd}).

\subsection{Carbon Stars}

Our observations in the two narrow band filters allow us to identify
carbon stars in the field of \andix. The survey technique described in
detail in \citet{harbeck2004}, where we searched for carbon stars as
tracers of intermediate-age (1--10 Gyr) stellar populations in the
M\,31 dSph satellites And\,III, And\,V, And\,VI and And\,VII. With the
same selection criteria as given in our earlier paper we are able to
identify one carbon star candidate in the sight line of \andix
(plotted with a filled circle in Fig\,\ref{fig_cmd}). However, the
proximity of M\,31 makes membership confirmation of this carbon star
by its radial velocity essential for further conclusions, since it
could also be a member of M\,31's extended disk population (see
\citealt{battinelli2003}). We conclude here that -- based on the C
star survey -- there is no evidence for a substantial intermediate-age
stellar population in \andix, which is further corroborated by the
apparent absence of any noticeable number of luminous asymptotic
branch stars above the TRGB.  However, due to the small number of
stars in And IX, the statistical reliability of this method is
reduced. In any case we do not expect a substantial contamination of
the TRGB region by younger RGB stars, further improving the
reliability of the TRGB distance derived in Sect. 3.1.

\subsection{Spatial distribution and stellar density profile}
\begin{figure}
\plotone{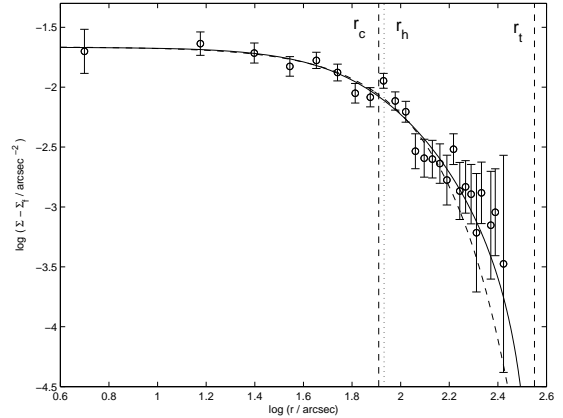}
\caption{Background-corrected surface density of \andix RGB stars
  plotted versus the radial distance to the center. Both a King
  profile (solid line) and a S\'ersic profile (dashed line) fit the
  data well (the data is actually fitted for $\log r \le 2.2$ only) .
  The King fit yields a core radius of $r_c = 1.33'$ and a tidal
  radius of $r_t = 5.9'$. The core and tidal radii are indicated by
  vertical lines. We derive a half light radius of $r_h = 85.5''$ from
  the S\'ersic profile.
\label{fig_spatial}}
\end{figure}

\begin{figure}
\plotone{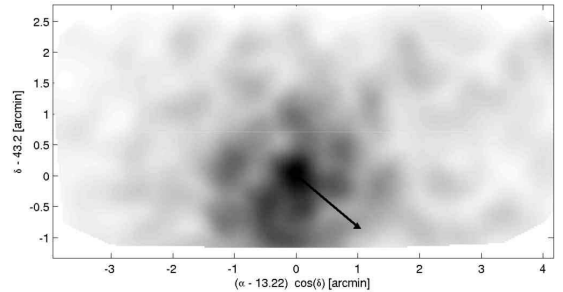}
\caption{Two-dimensional surface density of RGB stars in the \andix
  field, which shows signs of a clumpy structure, which is most likely
  attributed to the sparse sampling. The arrow indicates the direction
  towards the center of M\,31; no obvious elongation of \andix in this
  direction is apparent.
\label{fig_density}}
\end{figure}

The two dimensional density distribution of \andix's RGB stars
(Fig.~\ref{fig_density}) shows a very clumpy structure, which is
likely attributed to the scarcity of stars along the RGB. There is no
evidence for an elongation in the direction of M31, which might be
expected if tidal forces were at work.

We present the stellar density profile for stars in the RGB region of
the \andix CMD in Fig.~\ref{fig_spatial}, assuming a radially
symmetric distribution; the sparse sampling does not permit us to
derive further parameters such as ellipticity or position angle, and
we use the same center as in \citet{zucker2004}. We use a theoretical
King profile \citep{king1962} as well as a general exponential
S\'ersic profile \citep{sersic1968} to parameterize the radial density
distribution (Fig.~\ref{fig_spatial}). We find the core radius ($r_c$)
and tidal radius ($r_t$) of \andix to be $1.35'$ and $5.9'$,
respectively, for the King fit. The S\'ersic profile results in a half
light radius of $1.43'$ and an exponent of 1.6. The half-light radius
is in accordance with the one used in \citet{huxor2004}. In their
Fig.\,7, \andix half-light radius match those of other dSphs, too, in
contrast to star clusters. As our study still suffers from low number
statistics, future deeper studies will be able to increase the
significance of the structural parameters.

\subsection{Global properties}
\begin{figure}
\plotone{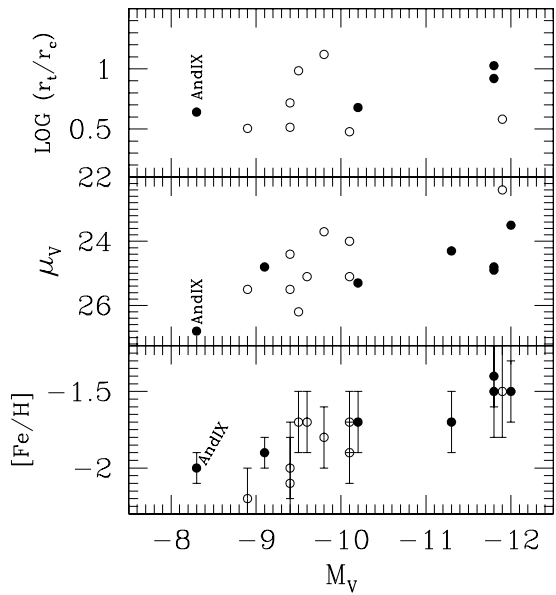}
\caption{Luminosity-metallicity relation for Local Group dSph
  satellite subsystem (lower panel). Error bars represent the range of
  metallicities in a galaxy. \andix, the least massive dSph in this
  system, clearly fits into the relation, which argues that \andix is
  a primordial low-mass dwarf galaxy rather than tidal debris of a
  more massive dwarf galaxy. The data are from \citet{grebel2003},
  \citet{zucker2004}, and this work. The luminosity surface brightness
  relation of LG dSph galaxies is shown in the middle panel. While
  \andix has a very low surface brightness, within the scatter it fits
  well into the general trend of the other dSphs. Filled circles refer
  to M\,31 dSphs, while open circles represent other LG dSph
  galaxies. The upper panel shows the ratio of the tidal radius to the
  core radius (data from compilation in \citealt{harbeck2001} and
  \citealt{irwin1995}).
\label{fig_lvfeh}}
\end{figure}
The luminosity-metallicity relation of LG dSph galaxies is plotted in
Fig.~\ref{fig_lvfeh} (lower panel). Given its low luminosity, \andix
fits the general luminosity-metallicity relation that is followed by
dSph galaxies (Grebel et al.\ 2003).  In Fig.~\ref{fig_lvfeh} (middle
panel) we plot the luminosity against the surface brightness; although
the surface brightness of \andix is remarkably low, it appears to
follow the general trend of LG galaxies. The ratio of the tidal radius
to the core radius ($r_t \over r_c$) is well within the range of other
LG dSph galaxies (Fig.~\ref{fig_lvfeh} upper panel). Apart from its
low luminosity, \andix compares well to other LG dSph galaxies.

\section{Conclusions}

Our WIYN observations confirm that \andix is a dSph companion of M\,31
as previously claimed by \citet{zucker2004} and
\citet{mcconnachie2004} on the basis of shallower observations. Its
low metallicity is consistent with the low luminosity of this galaxy
and fits well into the M$_V$--[Fe/H] relation of the LG dSph
system. We are not able to detect a significant population of carbon
stars in \andix, nor do we see evidence for a bright main sequence,
extended AGB, or a young red clump in the color magnitude diagram,
which suggests that there is no substantial population of
intermediate-age stars present. Our observations imply that \andix is
indeed an old, sparse, very metal-deficient dSph galaxy.

Ancient dwarf galaxies such as \andix with extremely low baryonic
masses are of special interest in constraining theories of galaxy
formation. For example, in models with strong feedback or severe
repression of early gas accretion during reionization, such galaxies
are difficult to form (e.g., \citealt{thoul1996},
\citealt{shapiro2004}). An alternative approach produces low baryon
mass galaxies via interactions including stripping of what initially
were larger objects (\citealt{mayer2001a,mayer2001b,kravtsov2004} and
references therein). The properties of \andix, the least massive of
M31's satellites, allows us to place additional constraints on such
models.

A key factor is the apparently normal position of \andix in the
luminosity-metallicity and luminosity-$\mu_V$ diagrams for dSph
galaxies. This precludes its being a product of a galaxy that once had
a significantly larger stellar mass.  Observations show that more
massive dSph systems have higher metallicities, even when, as in the
M31 satellites, they consist almost entirely of stars older than
8--10~Gyr (e.g., \citealt{harbeck2001, dacosta2002, harbeck2004}). If
small galaxies are simply the central remnants of larger objects then
they should have a metallicity appropriate to their state when most of
their stars formed. Thus if the stripping occurred after more than 3-4
Gyr following formation, when the M31 dSphs were nearly chemical
mature, then the metallicity of a remnant such as \andix should be
higher than expected for its present day luminosity.

Making a low metallicity galaxy like \andix from a more massive object
therefore requires that the baryon reservoir is removed at an early
time, before the galaxy could chemically enrich itself (see Grebel \&
Gallagher 2004 and references therein). As a result the ratio of
baryonic to dark matter masses could be well below the cosmic value.
Unfortunately, the tidal stripping models of \citet{kravtsov2004}
require 3-4 Gyr for large amounts of stellar mass to be stripped, and
none of these models reach the stellar mass level of \andix. Possibly
ram pressure stripping of the ISM played an equally important role, as
suggested more generally for the production of dSphs by
\citet{grebel2003}. The present day proximity of \andix to M\,31
suggests it is in a close orbit, which might argue for enhanced gas
stripping in its early evolution.

Of course And IX may simply be an intrinsically low mass galaxy. In
this case the issue becomes balancing the ease of gas loss, especially
when near a giant galaxy that is forming, against the need to form
some stars. The possibility that even very low mass galaxies may be
able to make stars if they collapse at sufficiently high redshifts
\citep{barkana1999, dijkstra2004, grebel2004} then is a critical
factor.  Furthermore, low mass galaxies are the most likely survivors
on close orbits since they experience less dynamical friction (e.g.,
\citealt{taylor2001}).

In summary, And IX appears to be a low mass but otherwise normal dSph
galaxy. The detection of one additional dSph satellite of M31 does not
in itself solve the problem of missing satellites in galaxy halos
\citep{moore1999}. However, the presence of a system with such a low
stellar density in the proximity of M31 raises the possibility that
dark halos may exist with very low stellar masses which could be
missed even in existing surveys.

\acknowledgments

We would like to thank the WIYN staff for their great support. DH
gratefully acknowledges support as a McKinney Fellow at the University
of Wisconsin, Madison, and as a Townes Fellow at the University of
California, Berkeley. JSG is supported by NSF grant AST98-03018 and
the Wisconsin Alumni Research Foundation, and EKG and AK by the Swiss
National Science Foundation through grant 200021-101924/1.  DBZ
received partial support from a National Science Foundation
International Postdoctoral Fellowship. We thank the referee for
his/her helpful comments on the manuscript.

This research has made use of the NASA/IPAC Extragalactic Database
(NED) which is operated by the Jet Propulsion Laboratory, California
Institute of Technology, under contract with the National Aeronautics
and Space Administration.




\end{document}